\documentclass[a4paper,11pt]{article}

\makeatletter

\gdef\@fpheader{\newline}
\gdef\@journal{jhep}

\RequirePackage{amsmath}
\RequirePackage{amssymb}
\RequirePackage{epsfig}
\RequirePackage{CJK}
\RequirePackage{graphicx}
\RequirePackage[numbers,sort&compress]{natbib}
\RequirePackage{color}
\RequirePackage[colorlinks=true
,urlcolor=blue
,anchorcolor=blue
,citecolor=blue
,filecolor=blue
,linkcolor=blue
,menucolor=blue
,pagecolor=blue
,linktocpage=true
,pdfproducer=medialab
,pdfa=true
,CJKbookmarks
]{hyperref}

\newif\ifnotoc\notocfalse
\newif\ifemailadd\emailaddfalse
\newif\iftoccontinuous\toccontinuousfalse

\def\@subheader{\@empty}
\def\@keywords{\@empty}
\def\@abstract{\@empty}
\def\@xtum{\@empty}
\def\@dedicated{\@empty}
\def\@arxivnumber{\@empty}
\def\@collaboration{\@empty}
\def\@collaborationImg{\@empty}
\def\@proceeding{\@empty}
\def\@preprint{\@empty}

\newcommand{\subheader}[1]{\gdef\@subheader{#1}}
\newcommand{\keywords}[1]{\if!\@keywords!\gdef\@keywords{#1}\else%
\PackageWarningNoLine{\jname}{Keywords already defined.\MessageBreak Ignoring last definition.}\fi}
\renewcommand{\abstract}[1]{\gdef\@abstract{#1}}
\newcommand{\dedicated}[1]{\gdef\@dedicated{#1}}
\newcommand{\arxivnumber}[1]{\gdef\@arxivnumber{#1}}
\newcommand{\proceeding}[1]{\gdef\@proceeding{#1}}
\newcommand{\xtumfont}[1]{\textsc{#1}}
\newcommand{\correctionref}[3]{\gdef\@xtum{\xtumfont{#1} \href{#2}{#3}}}
\newcommand\jname{JHEP}
\newcommand\acknowledgments{\section*{Acknowledgments}}

\newcommand\preprint[1]{\gdef\@preprint{\hfill #1}}

\newcommand\note[2][]{%
\if!#1!%
\stepcounter{footnote}\footnotetext{#2}%
\else%
{\renewcommand\thefootnote{#1}%
\footnotetext{#2}}%
\fi}

\newtoks\auth@toks
\renewcommand{\author}[2][]{%
  \if!#1!%
    \auth@toks=\expandafter{\the\auth@toks#2\ }%
  \else
    \auth@toks=\expandafter{\the\auth@toks#2$^{#1}$\ }%
  \fi
}

\newtoks\affil@toks\newif\ifaffil\affilfalse
\newcommand{\affiliation}[2][]{%
\affiltrue
  \if!#1!%
    \affil@toks=\expandafter{\the\affil@toks{\item[]#2}}%
  \else
    \affil@toks=\expandafter{\the\affil@toks{\item[$^{#1}$]#2}}%
  \fi
}

\newtoks\email@toks\newcounter{email@counter}%
\newcommand{\emailAdd}[1]{%
\emailaddtrue%
\ifnum\theemail@counter>0\email@toks=\expandafter{\the\email@toks, \@email{#1}}%
\else\email@toks=\expandafter{\the\email@toks\@email{#1}}%
\fi\stepcounter{email@counter}}
\newcommand{\@email}[1]{\href{mailto:#1}{\tt #1}}

\newcommand*\collaboration[1]{\gdef\@collaboration{#1}}
\newcommand*\collaborationImg[2][]{\gdef\@collaborationImg{#2}}

\newcommand\afterLogoSpace{\smallskip}
\newcommand\afterSubheaderSpace{\vskip3pt plus 2pt minus 1pt}
\newcommand\afterProceedingsSpace{\vskip21pt plus0.4fil minus15pt}
\newcommand\afterTitleSpace{\vskip23pt plus0.06fil minus13pt}
\newcommand\afterRuleSpace{\vskip23pt plus0.06fil minus13pt}
\newcommand\afterCollaborationSpace{\vskip3pt plus 2pt minus 1pt}
\newcommand\afterCollaborationImgSpace{\vskip3pt plus 2pt minus 1pt}
\newcommand\afterAuthorSpace{\vskip5pt plus4pt minus4pt}
\newcommand\afterAffiliationSpace{\vskip3pt plus3pt}
\newcommand\afterEmailSpace{\vskip16pt plus9pt minus10pt\filbreak}
\newcommand\afterXtumSpace{\par\bigskip}
\newcommand\afterAbstractSpace{\vskip16pt plus9pt minus13pt}
\newcommand\afterKeywordsSpace{\vskip16pt plus9pt minus13pt}
\newcommand\afterArxivSpace{\vskip3pt plus0.01fil minus10pt}
\newcommand\afterDedicatedSpace{\vskip0pt plus0.01fil}
\newcommand\afterTocSpace{\bigskip\medskip}
\newcommand\afterTocRuleSpace{\bigskip\bigskip}
\newlength{\affiliationsSep}
\newcommand\beforetochook{\pagestyle{myplain}\pagenumbering{roman}}

\DeclareFixedFont\trfont{OT1}{phv}{b}{sc}{11}

\renewcommand\maketitle{
\pagestyle{empty}
\thispagestyle{titlepage}
\setcounter{page}{0}
\noindent{\small\scshape\@fpheader}\@preprint\par
\afterLogoSpace
\if!\@subheader!\else\noindent{\trfont{\@subheader}}\fi
\afterSubheaderSpace
\if!\@proceeding!\else\noindent{\sc\@proceeding}\fi
\afterProceedingsSpace
{\LARGE\flushleft\sffamily\bfseries\@title\par}
\afterTitleSpace
\hrule height 1.5\p@%
\afterRuleSpace
\if!\@collaboration!\else
{\Large\bfseries\sffamily\raggedright\@collaboration}\par
\afterCollaborationSpace
\fi
\if!\@collaborationImg!\else
{\normalsize\bfseries\sffamily\raggedright\@collaborationImg}\par
\afterCollaborationImgSpace
\fi
{\bfseries\raggedright\sffamily\the\auth@toks\par}
\afterAuthorSpace
\ifaffil\begin{list}{}{%
\setlength{\leftmargin}{0.28cm}%
\setlength{\labelsep}{0pt}%
\setlength{\itemsep}{\affiliationsSep}%
\setlength{\topsep}{-\parskip}}
\itshape\small%
\the\affil@toks
\end{list}\fi
\afterAffiliationSpace
\ifemailadd 
\noindent\hspace{0.28cm}\begin{minipage}[l]{.9\textwidth}
\begin{flushleft}
\textit{E-mail:} \the\email@toks
\end{flushleft}
\end{minipage}
\else 
\PackageWarningNoLine{\jname}{E-mails are missing.\MessageBreak Plese use \protect\emailAdd\space macro to provide e-mails.}
\fi
\afterEmailSpace
\if!\@xtum!\else\noindent{\@xtum}\afterXtumSpace\fi
\if!\@abstract!\else\noindent{\renewcommand\baselinestretch{.9}\textsc{Abstract:}}\ \@abstract\afterAbstractSpace\fi
\if!\@keywords!\else\noindent{\textsc{Keywords:}} \@keywords\afterKeywordsSpace\fi
\if!\@arxivnumber!\else\noindent{\textsc{ArXiv ePrint:}} \href{http://arxiv.org/abs/\@arxivnumber}{\@arxivnumber}\afterArxivSpace\fi
\if!\@dedicated!\else\vbox{\small\it\raggedleft\@dedicated}\afterDedicatedSpace\fi
\ifnotoc\else
\iftoccontinuous\else\newpage\fi
\beforetochook\hrule
\tableofcontents
\afterTocSpace
\hrule
\afterTocRuleSpace
\fi
\setcounter{footnote}{0}
\pagestyle{myplain}\pagenumbering{arabic}
} 

\renewcommand{\baselinestretch}{1.1}\normalsize
\divide\@tempdima\baselineskip
\@tempcnta=\@tempdima
\skip\@mpfootins = \skip\footins
\renewcommand{\@dotsep}{10000}

\newcommand\ps@myplain{
\pagenumbering{arabic}
\renewcommand\@oddfoot{\hfill-- \thepage\ --\hfill}
\renewcommand\@oddhead{}}
\let\ps@plain=\ps@myplain

\newcommand\ps@titlepage{\renewcommand\@oddfoot{}\renewcommand\@oddhead{}}

\numberwithin{equation}{section}

\renewcommand\section{\@startsection{section}{1}{\z@}%
                                   {-3.5ex \@plus -1.3ex \@minus -.7ex}%
                                   {2.3ex \@plus.4ex \@minus .4ex}%
                                   {\normalfont\large\bfseries}}
\renewcommand\subsection{\@startsection{subsection}{2}{\z@}%
                                   {-2.3ex\@plus -1ex \@minus -.5ex}%
                                   {1.2ex \@plus .3ex \@minus .3ex}%
                                   {\normalfont\normalsize\bfseries}}
\renewcommand\subsubsection{\@startsection{subsubsection}{3}{\z@}%
                                   {-2.3ex\@plus -1ex \@minus -.5ex}%
                                   {1ex \@plus .2ex \@minus .2ex}%
                                   {\normalfont\normalsize\bfseries}}
\renewcommand\paragraph{\@startsection{paragraph}{4}{\z@}%
                                   {1.75ex \@plus1ex \@minus.2ex}%
                                   {-1em}%
                                   {\normalfont\normalsize\bfseries}}
\renewcommand\subparagraph{\@startsection{subparagraph}{5}{\parindent}%
                                   {1.75ex \@plus1ex \@minus .2ex}%
                                   {-1em}%
                                   {\normalfont\normalsize\bfseries}}

\def\fnum@figure{\textbf{\figurename\nobreakspace\thefigure}}
\def\fnum@table{\textbf{\tablename\nobreakspace\thetable}}

\long\def\@makecaption#1#2{%
  \vskip\abovecaptionskip
  \sbox\@tempboxa{\small #1. #2}%
  \ifdim \wd\@tempboxa >\hsize
    \small #1. #2\par
  \else
    \global \@minipagefalse
    \hb@xt@\hsize{\hfil\box\@tempboxa\hfil}%
  \fi
  \vskip\belowcaptionskip}

\renewenvironment{thebibliography}[1]{%
\begin{oldthebibliography}{#1}%
\small%
\raggedright%
\setlength{\itemsep}{5pt plus 0.2ex minus 0.05ex}%
}%
{%
\end{oldthebibliography}%
}

\makeatother

\usepackage[T1]{fontenc} 
\usepackage{romannum}

\begin{document} 


\title{{\boldmath Cylindrical gravitational waves: radiation and resonance} \\  }

\author[a,1]{Yu-Zhu Chen,}\note{chenyuzhu@nankai.edu.cn}
\author[b]{Yu-Jie Chen,}
\author[b]{Shi-Lin Li,}
\author[b,2]{and Wu-Sheng Dai}\note{daiwusheng@tju.edu.cn.}


\affiliation[a]{Theoretical Physics Division, Chern Institute of Mathematics, Nankai University, Tianjin, 300071, P. R. China}
\affiliation[b]{Department of Physics, Tianjin University, Tianjin 300350, P.R. China}








\abstract{
In the weak field approximation the gravitational wave is approximated as a
linear wave, which ignores the nonlinear effect. In this paper, we present an
exact general solution of the cylindrical gravitational wave. The exact
solution of the cylindrical gravitational wave is far different from the weak
field approximation. This solution implies the following conclusions. (1)
There exist gravitational monopole radiations\textbf{ }in the cylindrical
gravitational radiation. (2) The gravitational radiation may generate the
resonance in the spacetime. (3) The nonlinearity of the gravity source
vanishes after time averaging, so the observed result of a long-time
measurement may be linear.

}




\maketitle 

\flushbottom

\section{Introduction}

General relativity is a nonlinear theory. The quadrupole radiation, the
polarization, the spin of the graviton, however, are mostly obtained under the
linear approximation \cite{weinberg1972gravitation,ohanian2013gravitation}. In
the linear approximation, it is supposed that the energy and the momentum of
the source are invariable. As a result, the monopole and the dipole of the
source are invariable and make no contribution to the gravitational radiation
\cite{misner2017gravitation}. Besides, when the spacetime is regarded as a
Minkowski spacetime, the dipole vanishes in the center-of-mass coordinate
\cite{ohanian2013gravitation}. Therefore, the leading contribution of the
gravitational radiation is made by the variation of the quadrupole
\cite{weinberg1972gravitation,ohanian2013gravitation}, i.e., the gravitational
quadrupole radiation.

In this paper, we present an exact cylindrical solution which implies a
gravitational monopole radiation. The exact nonlinear solution is far
different from the linear approximation. By the Birkhoff theorem we know that
there is no gravitational wave in the spherical vacuum solution of the
Einstein equation
\cite{weinberg1972gravitation,ohanian2013gravitation,misner2017gravitation}.
The simplest metric involving the gravitational radiation is the
Einstein-Rosen metric. The source of an infinite length may exist in the
Einstein-Rosen metric. It is convenient to discuss the gravitational radiation
with the Einstein-Rosen metric, such as the energy loss of the source by the
gravitational radiation \cite{thorne1965energy}, the C-energy, the
super-energy, and the associated dynamical effect of the cylindrical
gravitational wave \cite{bini2019cylindrical}, the energy-momentum of the
gravitational wave \cite{garecki2016gravitational}, the superenergy flux of
the Einstein-Rosen wave \cite{dominguez2018superenergy}, the nonlinear effect
such as the Faraday rotation and the time-shift phenomenon of the cylindrical
gravitational soliton solution \cite{tomizawa2015nonlinear}, the nonlinear
evolution of cylindrical gravitational waves \cite{celestino2016nonlinear},
the twisted gravitational wave \cite{bini2018twisted}, the scattering of the
gravitational waves \cite{chandrasekhar1987dispersion}, the gravitational
collapse of the energy of gravitational waves
\cite{chakraborty2016gravitational}\textbf{, }the asymptotic structure of the
radiation spacetime \cite{stachel1966cylindrical}, the interaction between the
gravitational wave and the cosmic string
\cite{garriga1987cosmic,xanthopoulos1987cosmic}, the cosmic censorship
hypothesis \cite{berger1995asymptotically}, and the midisuperspace
quantization
\cite{korotkin1998canonical,kuchavr1971canonical,ashtekar1996probing}.

We also discuss the resonance of the gravitational radiation in the spacetime.
In literature, the resonance between the gravitational radiation and the
matter (especially the gravitational radiation detectors) are considered
\cite{detweiler1980black,hopman2006resonant,mizuno1993resonant,staley2014achieving}%
. In this paper, we focus on the resonance of gravitational waves.

In the linear approximation, two gravitational radiations do not interact with
each other. In this paper, we show the interaction between two cylindrical
gravitational radiations. When two cylindrical gravitational radiations exist
simultaneously, the interaction terms arise both in the metric and the
energy-momentum tensor. Especially, we show that, though the interaction
always exists in nonlinear gravity waves, the nonlinearity of the gravity
source vanishes when taking a time average. This implies that the observed
result of a long-time measurement may be linear, though the gravity is nonlinear.

In section 2, we present an exact general cylindrical gravitational wave
solution. In section 3, we show the existence of the cylindrical gravitational
monopole radiation. In section 4, we discuss the resonance of the
gravitational radiation. In section 5, we show the interaction of two
cylindrical gravitational radiations. The conclusion and outlooks are given in
section 6.

\section{Cylindrical gravitational wave: general solution}

In this section, we present a general vacuum solution of the cylindrical
gravitational wave. With the cylindrical gravitational wave solution, we
discuss the gravitational monopole radiation, the resonance, and the
interaction of the cylindrical gravitational radiations in the latter sections.

The cylindrical gravitational wave is described by the $1+3$ dimensional
Einstein-Rosen metric \cite{carmeli2001classical},
\begin{equation}
ds^{2}=e^{2\gamma\left(  t,\rho\right)  -2\psi\left(  t,\rho\right)  }\left(
-dt^{2}+d\rho^{2}\right)  +e^{-2\psi\left(  t,\rho\right)  }\rho^{2}d\phi
^{2}+e^{2\psi\left(  t,\rho\right)  }dz^{2}. \label{Metric}%
\end{equation}
The Einstein tensor $G_{\mu\nu}=R_{\mu\nu}-\frac{1}{2}\eta_{\mu\nu}R$ in the
orthogonal frame \cite{chen2017dimensional} with $\eta_{\mu\nu}%
=\operatorname*{diag}\left(  -1,1,1,1\right)  $ is%
\begin{align}
G_{00}  &  =G_{11}=e^{2\psi-2\gamma}\left[  \frac{1}{\rho}\frac{\partial
\gamma}{\partial\rho}-\left(  \frac{\partial\psi}{\partial\rho}\right)
^{2}-\left(  \frac{\partial\psi}{\partial t}\right)  ^{2}\right]
,\label{T00}\\
G_{01}  &  =G_{10}=e^{2\psi-2\gamma}\left(  \frac{1}{\rho}\frac{\partial
\gamma}{\partial t}-2\frac{\partial\psi}{\partial t}\frac{\partial\psi
}{\partial\rho}\right)  ,\label{T01}\\
G_{22}  &  =e^{2\psi-2\gamma}\left[  \left(  \frac{\partial\psi}{\partial\rho
}\right)  ^{2}-\left(  \frac{\partial\psi}{\partial t}\right)  ^{2}%
+\frac{\partial^{2}\gamma}{\partial\rho^{2}}-\frac{\partial^{2}\gamma
}{\partial t^{2}}\right]  ,\label{T22}\\
G_{33}  &  =e^{2\psi-2\gamma}\left[  \left(  \frac{\partial\psi}{\partial\rho
}\right)  ^{2}-\left(  \frac{\partial\psi}{\partial t}\right)  ^{2}%
+\frac{\partial^{2}\gamma}{\partial\rho^{2}}-\frac{\partial^{2}\gamma
}{\partial t^{2}}+2\frac{\partial^{2}\psi}{\partial t^{2}}-2\frac{\partial
^{2}\psi}{\partial\rho^{2}}-\frac{2}{\rho}\frac{\partial\psi}{\partial\rho
}\right]  . \label{T33}%
\end{align}

We consider the gravitational field outside the source, i.e., $\rho\neq0$,
which satisfies $G_{\mu\nu}=0$. For $G_{\mu\nu}=0$, eqs. (\ref{T00}),
(\ref{T01}), (\ref{T22}), and (\ref{T33}) can be simplified as
\cite{carmeli2001classical}%
\begin{align}
\frac{\partial\gamma}{\partial\rho}  &  =\rho\left(  \frac{\partial\psi
}{\partial t}\right)  ^{2}+\rho\left(  \frac{\partial\psi}{\partial\rho
}\right)  ^{2},\label{Eq.1}\\
\frac{\partial\gamma}{\partial t}  &  =2\rho\frac{\partial\psi}{\partial
t}\frac{\partial\psi}{\partial\rho}. \label{Eq.2}%
\end{align}
For gravitational wave solutions, we require $\frac{\partial\psi}{\partial
t}\neq0$. Then we obtain the equation of $\psi$ from eqs. (\ref{Eq.1}) and
(\ref{Eq.2}) \cite{carmeli2001classical},%
\begin{equation}
-\frac{\partial^{2}\psi}{\partial t^{2}}+\frac{\partial^{2}\psi}{\partial
\rho^{2}}+\frac{1}{\rho}\frac{\partial\psi}{\partial\rho}=0. \label{Eq.3}%
\end{equation}
The equation of $\psi$ (\ref{Eq.3}) is a linear equation. The general solution
of eq. (\ref{Eq.3}) is%

\begin{align}
\psi &  =\int_{-\infty}^{\infty}d\tau\int_{0}^{\infty}d\lambda A\left(
\tau,\lambda\right)  J_{0}\left(  \lambda\rho\right)  \cos\left(
\lambda\left(  t-\tau\right)  +\alpha\left(  \tau,\lambda\right)  \right)
\nonumber\\
&  +\int_{-\infty}^{\infty}d\tau\int_{0}^{\infty}d\lambda B\left(
\tau,\lambda\right)  Y_{0}\left(  \lambda\rho\right)  \cos\left(
\lambda\left(  t-\tau\right)  +\beta\left(  \tau,\lambda\right)  \right)
\nonumber\\
&  +\kappa_{1}t+\kappa_{2}\ln\rho+\kappa_{0}, \label{ps.general}%
\end{align}
where $\alpha\left(  \tau,\lambda\right)  $, $\beta\left(  \tau,\lambda
\right)  $, $A\left(  \tau,\lambda\right)  $, and $B\left(  \tau
,\lambda\right)  $ are arbitrary functions of $\tau$ and $\lambda$, $J_{0}$ is
the Bessel function of first kind, $Y_{0}$ is the Bessel function of second
kind, and $\kappa_{0}$, $\kappa_{1}$, and $\kappa_{2}$ are constants. From
eqs. (\ref{Eq.1}) and (\ref{Eq.2}) we can see that the equations of $\gamma$
are also linear equations. Substituting eq. (\ref{ps.general}) into eqs.
(\ref{Eq.1}) and (\ref{Eq.2}) gives the general solution of $\gamma$:
\begin{align}
\gamma &  =-\frac{\rho}{2}\int_{-\infty}^{\infty}d\tau_{1}\int_{-\infty
}^{\infty}d\tau_{2}\int_{0}^{\infty}d\lambda_{1}\int_{0}^{\infty}d\lambda
_{2}(F_{+}+F_{-})\nonumber\\
&  -2\kappa_{1}\rho\int_{-\infty}^{\infty}d\tau\int_{0}^{\infty}%
d\lambda\left[  A\left(  \tau,\lambda\right)  J_{1}\left(  \lambda\rho\right)
\sin\left(  \lambda\left(  t-\tau\right)  +\alpha\left(  \tau,\lambda\right)
\right)  +B\left(  \tau,\lambda\right)  Y_{1}\left(  \lambda\rho\right)
\sin\left(  \lambda\left(  t-\tau\right)  +\beta\left(  \tau,\lambda\right)
\right)  \right] \nonumber\\
&  -2\kappa_{2}\int_{-\infty}^{\infty}d\tau\int_{0}^{\infty}d\lambda\left[
A\left(  \tau,\lambda\right)  J_{1}\left(  \lambda\rho\right)  \cos\left(
\lambda\left(  t-\tau\right)  +\alpha\left(  \tau,\lambda\right)  \right)
+B\left(  \tau,\lambda\right)  Y_{1}\left(  \lambda\rho\right)  \cos\left(
\lambda\left(  t-\tau\right)  +\beta\left(  \tau,\lambda\right)  \right)
\right] \nonumber\\
&  +2\kappa_{1}\kappa_{2}t+\frac{1}{2}\kappa_{1}^{2}\rho^{2}+\kappa_{2}^{2}%
\ln\rho+\kappa_{3} \label{gamma.general}%
\end{align}
with
\begin{align}
F_{+}  &  =\frac{\lambda_{1}\lambda_{2}}{\lambda_{1}+\lambda_{2}}\{A\left(
\tau_{1},\lambda_{1}\right)  A\left(  \tau_{2},\lambda_{2}\right)  \left[
J_{0}\left(  \lambda_{1}\rho\right)  J_{1}\left(  \lambda_{2}\rho\right)
+J_{0}\left(  \lambda_{2}\rho\right)  J_{1}\left(  \lambda_{1}\rho\right)
\right] \nonumber\\
&  \times\cos\left(  \lambda_{1}t+\lambda_{2}t-\lambda_{1}\tau_{1}-\lambda
_{2}\tau_{2}+\alpha\left(  \tau_{1},\lambda_{1}\right)  +\alpha\left(
\tau_{2},\lambda_{2}\right)  \right) \nonumber\\
&  +A\left(  \tau_{1},\lambda_{1}\right)  B\left(  \tau_{2},\lambda
_{2}\right)  \left[  J_{0}\left(  \lambda_{1}\rho\right)  Y_{1}\left(
\lambda_{2}\rho\right)  +Y_{0}\left(  \lambda_{2}\rho\right)  J_{1}\left(
\lambda_{1}\rho\right)  \right] \nonumber\\
&  \times\cos\left(  \lambda_{1}t+\lambda_{2}t-\lambda_{1}\tau_{1}-\lambda
_{2}\tau_{2}+\alpha\left(  \tau_{1},\lambda_{1}\right)  +\beta\left(  \tau
_{2},\lambda_{2}\right)  \right) \nonumber\\
&  +B\left(  \tau_{1},\lambda_{1}\right)  A\left(  \tau_{2},\lambda
_{2}\right)  \left[  Y_{0}\left(  \lambda_{1}\rho\right)  J_{1}\left(
\lambda_{2}\rho\right)  +J_{0}\left(  \lambda_{2}\rho\right)  Y_{1}\left(
\lambda_{1}\rho\right)  \right] \nonumber\\
&  \times\cos\left(  \lambda_{1}t+\lambda_{2}t-\lambda_{1}\tau_{1}-\lambda
_{2}\tau_{2}+\beta\left(  \tau_{1},\lambda_{1}\right)  +\alpha\left(  \tau
_{2},\lambda_{2}\right)  \right) \nonumber\\
&  +B\left(  \tau_{1},\lambda_{1}\right)  B\left(  \tau_{2},\lambda
_{2}\right)  \left[  Y_{0}\left(  \lambda_{1}\rho\right)  Y_{1}\left(
\lambda_{2}\rho\right)  +Y_{0}\left(  \lambda_{2}\rho\right)  Y_{1}\left(
\lambda_{1}\rho\right)  \right] \nonumber\\
&  \times\cos\left(  \lambda_{1}t+\lambda_{2}t-\lambda_{1}\tau_{1}-\lambda
_{2}\tau_{2}+\beta\left(  \tau_{1},\lambda_{1}\right)  +\beta\left(  \tau
_{2},\lambda_{2}\right)  \right)  ]
\end{align}
and
\begin{align}
F_{-}  &  =\frac{\lambda_{1}\lambda_{2}}{\lambda_{1}-\lambda_{2}}\{A\left(
\tau_{1},\lambda_{1}\right)  A\left(  \tau_{2},\lambda_{2}\right)  \left[
J_{0}\left(  \lambda_{1}\rho\right)  J_{1}\left(  \lambda_{2}\rho\right)
-J_{0}\left(  \lambda_{2}\rho\right)  J_{1}\left(  \lambda_{1}\rho\right)
\right] \nonumber\\
&  \times\cos\left(  \lambda_{1}t-\lambda_{2}t-\lambda_{1}\tau_{1}+\lambda
_{2}\tau_{2}+\alpha\left(  \tau_{1},\lambda_{1}\right)  -\alpha\left(
\tau_{2},\lambda_{2}\right)  \right) \nonumber\\
&  +A\left(  \tau_{1},\lambda_{1}\right)  B\left(  \tau_{2},\lambda
_{2}\right)  \left[  J_{0}\left(  \lambda_{1}\rho\right)  Y_{1}\left(
\lambda_{2}\rho\right)  -Y_{0}\left(  \lambda_{2}\rho\right)  J_{1}\left(
\lambda_{1}\rho\right)  \right] \nonumber\\
&  \times\cos\left(  \lambda_{1}t-\lambda_{2}t-\lambda_{1}\tau_{1}+\lambda
_{2}\tau_{2}+\alpha\left(  \tau_{1},\lambda_{1}\right)  -\beta\left(  \tau
_{2},\lambda_{2}\right)  \right) \nonumber\\
&  +B\left(  \tau_{1},\lambda_{1}\right)  A\left(  \tau_{2},\lambda
_{2}\right)  \left[  Y_{0}\left(  \lambda_{1}\rho\right)  J_{1}\left(
\lambda_{2}\rho\right)  -J_{0}\left(  \lambda_{2}\rho\right)  Y_{1}\left(
\lambda_{1}\rho\right)  \right] \nonumber\\
&  \times\cos\left(  \lambda_{1}t-\lambda_{2}t-\lambda_{1}\tau_{1}+\lambda
_{2}\tau_{2}+\beta\left(  \tau_{1},\lambda_{1}\right)  -\alpha\left(  \tau
_{2},\lambda_{2}\right)  \right) \nonumber\\
&  +B\left(  \tau_{1},\lambda_{1}\right)  B\left(  \tau_{2},\lambda
_{2}\right)  \left[  Y_{0}\left(  \lambda_{1}\rho\right)  Y_{1}\left(
\lambda_{2}\rho\right)  -Y_{0}\left(  \lambda_{2}\rho\right)  Y_{1}\left(
\lambda_{1}\rho\right)  \right] \nonumber\\
&  \times\cos\left(  \lambda_{1}t-\lambda_{2}t-\lambda_{1}\tau_{1}+\lambda
_{2}\tau_{2}+\beta\left(  \tau_{1},\lambda_{1}\right)  -\beta\left(  \tau
_{2},\lambda_{2}\right)  \right)  \}.
\end{align}
Here $\kappa_{3}$ in eq. (\ref{gamma.general}) can be eliminated by a
coordinate transformation and, without loss of generality, we set $\kappa
_{3}=0$.

The solutions (\ref{ps.general}) and (\ref{gamma.general}) are the general
solutions of the cylindrical gravitational wave. The solutions given in ref.
\cite{carmeli2001classical} and the solutions mentioned after are particular
cases of the solutions (\ref{ps.general}) and (\ref{gamma.general}).

\section{Gravitational monopole radiation}

In this section, we give an exact cylindrical gravitational monopole radiation
solution. The radiation is a wave produced by a source. For example, the plane
electromagnetic wave is not an electromagnetic radiation and thee
electromagnetic wave produced by the antenna is an electromagnetic radiation.
The gravitational radiation can be recognized by observing if a wave solution
has a time-varying source.

Next (1) we first choose a particular solution and show that the solution is a
gravitational wave solution. (2) We separate the monopole radiation solution
from this gravitational wave solution.

\subsection{Gravitational wave solution}

In this section, we present an exact solution, and we show that the solution
is a gravitational wave by investigating the Weinberg energy-momentum
pseudotensor \cite{weinberg1972gravitation} and the Laudau-Lifscitz
energy-momentum pseudotensor \cite{landau2013classical}. The energy-momentum
pseudotensors are defined to describe the energy-momentum of the gravitational
field. If the energy-momentum pseudotensor of the gravitational field is
time-varying, we say that the solution represents a gravitational wave
\cite{landau2013classical}.

Taking
\begin{align}
A\left(  \tau,\lambda\right)   &  =A\delta\left(  \omega-\lambda\right)
\delta\left(  \tau\right)  ,\nonumber\\
B\left(  \tau,\lambda\right)   &  =B\delta\left(  \omega-\lambda\right)
\delta\left(  \tau\right)  ,\nonumber\\
\kappa_{0}  &  =\kappa_{1}=\kappa_{2}=0
\end{align}
in eqs. \textbf{(}\ref{ps.general}\textbf{)} and \textbf{(}\ref{gamma.general}%
\textbf{)}, where $\omega\neq0$ is a constant, we obtain a particular
solution
\begin{align}
\psi &  =AJ_{0}\left(  \omega\rho\right)  \cos\left(  \omega t+\alpha\right)
+BY_{0}\left(  \omega\rho\right)  \cos\left(  \omega t+\beta\right)
,\label{single.ps}\\
\gamma &  =f\left(  \rho\right)  -\frac{2AB}{\pi}\omega t\sin\left(
\alpha-\beta\right) \nonumber\\
&  -\frac{A^{2}}{2}\omega\rho J_{0}\left(  \omega\rho\right)  J_{1}\left(
\omega\rho\right)  \cos\left(  2\omega t+2\alpha\right)  -\frac{B^{2}}%
{2}\omega\rho Y_{0}\left(  \omega\rho\right)  Y_{1}\left(  \omega\rho\right)
\cos\left(  2\omega t+2\beta\right) \nonumber\\
&  -\frac{AB}{2}\omega\rho\left[  J_{1}\left(  \omega\rho\right)  Y_{0}\left(
\omega\rho\right)  +J_{0}\left(  \omega\rho\right)  Y_{1}\left(  \omega
\rho\right)  \right]  \cos\left(  2\omega t+\alpha+\beta\right)
\label{single.gamma}%
\end{align}
with $\alpha\equiv\alpha\left(  0,\omega\right)  $, $\beta\equiv\beta\left(
0,\omega\right)  $, and
\begin{align}
f\left(  \rho\right)   &  =\frac{A^{2}}{4}\omega^{2}\rho^{2}\left[  J_{0}%
^{2}\left(  \omega\rho\right)  +2J_{1}^{2}\left(  \omega\rho\right)
-J_{0}\left(  \omega\rho\right)  J_{2}\left(  \omega\rho\right)  \right]
\nonumber\\
&  +\frac{B^{2}}{4}\omega^{2}\rho^{2}\left[  Y_{0}^{2}\left(  \omega
\rho\right)  +2Y_{1}^{2}\left(  \omega\rho\right)  -Y_{0}\left(  \omega
\rho\right)  Y_{2}\left(  \omega\rho\right)  \right] \nonumber\\
&  +\frac{AB}{4}\omega^{2}\rho^{2}\left[  2J_{0}\left(  \omega\rho\right)
Y_{0}\left(  \omega\rho\right)  +4J_{1}\left(  \omega\rho\right)  Y_{1}\left(
\omega\rho\right)  -J_{0}\left(  \omega\rho\right)  Y_{2}\left(  \omega
\rho\right)  -J_{2}\left(  \omega\rho\right)  Y_{0}\left(  \omega\rho\right)
\right]  \cos\left(  \alpha-\beta\right)  . \label{single.f}%
\end{align}
Note here that $\psi$ given by eq. (\ref{single.ps}) has been given in ref.
\cite{carmeli2001classical}; $\gamma$ given by eq. (\ref{single.gamma}) is
obtained in the present paper.

In the Cartesian coordinates\
\begin{align}
x  &  =\rho\cos\phi,\nonumber\\
y  &  =\rho\sin\phi,
\end{align}
the Einstein-Rosen metric reads
\begin{align}
ds^{2}  &  =-e^{2\gamma-2\psi}dt^{2}+\frac{e^{-2\psi}}{\rho^{2}}\left(
x^{2}e^{2\gamma}+y^{2}\right)  dx^{2}+\frac{e^{-2\psi}}{\rho^{2}}\left(
y^{2}e^{2\gamma}+x^{2}\right)  dy^{2}\nonumber\\
&  +\frac{2xy}{\rho^{2}}e^{-2\psi}\left(  e^{2\gamma}-1\right)  dxdy+e^{2\psi
}dz^{2}. \label{metric.Cartesian}%
\end{align}
It can be checked by the numerical method that the metric
(\ref{metric.Cartesian}) with $\psi$ given by eq. (\ref{single.ps}) and
$\gamma$ given by eq. (\ref{single.gamma}) has a time-varying Weinberg
energy-momentum pseudotensor \cite{weinberg1972gravitation} and a time-varying
Laudau-Lifscitz energy-momentum pseudotensor \cite{landau2013classical}.

Besides, there is a special case of the solutions (\ref{single.ps}) and
(\ref{single.gamma}) satisfying the out-going wave condition: when
$t-r=const$, $\psi$ and $\gamma$ remain unchanged at $\rho\rightarrow\infty$.
When
\begin{align}
A  &  =B,\nonumber\\
\beta &  =\alpha-\frac{\pi}{2},
\end{align}
the approximation of $\psi$ and $\gamma$ at $\rho\rightarrow\infty$ are
\begin{align}
\psi &  =A\sqrt{\frac{2}{\pi\omega\rho}}\cos\left(  \omega t-\omega\rho
+\alpha+\frac{1}{4}\pi\right)  ,\nonumber\\
\gamma &  =\frac{A^{2}}{\pi}\left[  \cos\left(  2\omega t-2\omega\rho
+2\alpha\right)  -2\left(  \omega t-\omega\rho\right)  \right]  .
\end{align}
In addition, it can be found that when the out-going wave condition is
satisfied, the curvature of the spacetime vanishes at $\rho\rightarrow\infty$.
That is, the spacetime is asymptotically flat in spite of that the metric is
not asymptotic to the Minkowski metric.

\subsection{Gravitational radiation}

In this section, we show that the solutions (\ref{single.ps}) and
(\ref{single.gamma}) contain a gravitational radiation. As mentioned above,
the gravitational radiation is a gravitational wave solution with a source.

The solutions (\ref{single.ps}) and (\ref{single.gamma}) have two special cases.

$A=0$:
\begin{align}
\psi_{\text{rad}}  &  =BY_{0}\left(  \omega\rho\right)  \cos\left(  \omega
t+\beta\right)  ,\label{ps.radiation}\\
\gamma_{\text{rad}}  &  =\frac{B^{2}}{4}\omega^{2}\rho^{2}\left[  Y_{0}%
^{2}\left(  \omega\rho\right)  +2Y_{1}^{2}\left(  \omega\rho\right)
-Y_{0}\left(  \omega\rho\right)  Y_{2}\left(  \omega\rho\right)  \right]
\nonumber\\
&  -\frac{B^{2}}{2}\omega\rho Y_{0}\left(  \omega\rho\right)  Y_{1}\left(
\omega\rho\right)  \cos\left(  2\omega t+2\beta\right)  ,
\label{gamma.radiation}%
\end{align}

$B=0$:
\begin{align}
\psi_{\text{nrad}}  &  =AJ_{0}\left(  \omega\rho\right)  \cos\left(  \omega
t+\alpha\right)  ,\label{ps.wave}\\
\gamma_{\text{nrad}}  &  =\frac{A^{2}}{4}\omega^{2}\rho^{2}\left[  J_{0}%
^{2}\left(  \omega\rho\right)  +2J_{1}^{2}\left(  \omega\rho\right)
-J_{0}\left(  \omega\rho\right)  J_{2}\left(  \omega\rho\right)  \right]
\nonumber\\
&  -\frac{A^{2}}{2}\omega\rho J_{0}\left(  \omega\rho\right)  J_{1}\left(
\omega\rho\right)  \cos\left(  2\omega t+2\alpha\right)  . \label{gamma.wave}%
\end{align}
These two cases, eqs. (\ref{ps.radiation}), (\ref{gamma.radiation}),
(\ref{ps.wave}), and (\ref{gamma.wave}), have been given in ref.
\cite{carmeli2001classical}.

In this paper, we point out that $\psi_{\text{rad}}$ and $\gamma_{\text{rad}}$
reprensent radiations, but $\psi_{\text{nrad}}$ and $\gamma_{\text{nrad}}$ do
not reprensent radiations. Here we use the subscripts "rad" to denote the
gravitational radiation and use "nrad" to denote the nonradiation
gravitational wave.

Next we show that $\psi_{\text{rad}}$ and $\gamma_{\text{rad}}$ describe a
gravitational monopole radiation, which has a time-varying energy density or a
monopole, and $\psi_{\text{nrad}}$ and $\gamma_{\text{nrad}}$ are pure
gravitational waves without sources.

The solutions (\ref{single.ps}) and (\ref{single.gamma}) represent a
gravitational wave in the vacuum for $\rho>0$. Nevertheless, when $\rho=0$,
there may exist a source. The singularity in the solutions (\ref{single.ps})
and (\ref{single.gamma}), according to ref. \cite{carmeli2001classical}, might
be interpreted as a matter presented along the $z$ axis. Just as the Coulomb
potential $1/r$, when $r\neq0$, the solution is a vacuum solution;
nevertheless, there is a point charge at $r=0$. We use a standard mathematical
analysis method to calculate the energy-momentum tensor at $\rho=0$. More
details of this method can be found in our previous work
\cite{chen2017singular}. Replacing $\rho$ by $\sqrt{\rho^{2}+\epsilon^{2}}$ in
eqs. (\ref{single.ps}) and (\ref{single.gamma}) and substituting the metric
(\ref{Metric}) into eqs. (\ref{T00}), (\ref{T01}), (\ref{T22}) and
(\ref{T33}), by the Einstein equation $G_{\mu\nu}=8\pi T_{\mu\nu}$, we arrive
at
\begin{align}
T_{00}\left(  \omega,\epsilon\right)   &  =T_{11}\left(  \omega,\epsilon
\right) \nonumber\\
&  =\frac{e^{2\psi-2\gamma}}{8\pi}\frac{\omega^{2}\epsilon^{2}}{\rho
^{2}+\epsilon^{2}}\left[  AJ_{1}\left(  \omega\sqrt{\rho^{2}+\epsilon^{2}%
}\right)  \cos\left(  \omega t+\alpha\right)  +BY_{1}\left(  \omega\sqrt
{\rho^{2}+\epsilon^{2}}\right)  \cos\left(  \omega t+\beta\right)  \right]
^{2}, \label{T00.exact}%
\end{align}%
\begin{align}
T_{01}\left(  \omega,\epsilon\right)   &  =\frac{e^{2\psi-2\gamma}}{8\pi}%
\frac{2\omega^{2}\epsilon^{2}}{\rho\sqrt{\rho^{2}+\epsilon^{2}}}\left[
AJ_{0}\left(  \omega\sqrt{\rho^{2}+\epsilon^{2}}\right)  \sin\left(  \omega
t+\alpha\right)  +BY_{0}\left(  \omega\sqrt{\rho^{2}+\epsilon^{2}}\right)
\sin\left(  \omega t+\beta\right)  \right] \nonumber\\
&  \times\left[  AJ_{1}\left(  \omega\sqrt{\rho^{2}+\epsilon^{2}}\right)
\cos\left(  \omega t+\alpha\right)  +BY_{1}\left(  \omega\sqrt{\rho
^{2}+\epsilon^{2}}\right)  \cos\left(  \omega t+\beta\right)  \right]  ,
\label{T01.exact}%
\end{align}%
\begin{align}
T_{22}\left(  \omega,\epsilon\right)   &  =\frac{e^{2\psi-2\gamma}}{8\pi}%
\frac{\omega^{2}\epsilon^{2}}{\rho^{2}+\epsilon^{2}}\left[  AJ_{1}\left(
\omega\sqrt{\rho^{2}+\epsilon^{2}}\right)  \cos\left(  \omega t+\alpha\right)
+BY_{1}\left(  \omega\sqrt{\rho^{2}+\epsilon^{2}}\right)  \cos\left(  \omega
t+\beta\right)  \right]  ^{2}\nonumber\\
&  -\frac{e^{2\psi-2\gamma}}{8\pi}\frac{2\omega^{3}\epsilon^{2}}{\sqrt
{\rho^{2}+\epsilon^{2}}}[A^{2}J_{0}\left(  \omega\sqrt{\rho^{2}+\epsilon^{2}%
}\right)  J_{1}\left(  \omega\sqrt{\rho^{2}+\epsilon^{2}}\right)  \cos\left(
2\omega t+2\alpha\right) \nonumber\\
&  +B^{2}Y_{0}\left(  \omega\sqrt{\rho^{2}+\epsilon^{2}}\right)  Y_{1}\left(
\omega\sqrt{\rho^{2}+\epsilon^{2}}\right)  \cos\left(  2\omega t+2\beta
\right)  ]\nonumber\\
&  -\frac{e^{2\psi-2\gamma}}{8\pi}\frac{2\omega^{3}\epsilon^{2}}{\sqrt
{\rho^{2}+\epsilon^{2}}}AB[J_{0}\left(  \omega\sqrt{\rho^{2}+\epsilon^{2}%
}\right)  Y_{1}\left(  \omega\sqrt{\rho^{2}+\epsilon^{2}}\right) \nonumber\\
&  +J_{1}\left(  \omega\sqrt{\rho^{2}+\epsilon^{2}}\right)  Y_{0}\left(
\omega\sqrt{\rho^{2}+\epsilon^{2}}\right)  ]\cos\left(  2\omega t+\alpha
+\beta\right)  , \label{T22.exact}%
\end{align}%
\begin{align}
T_{33}\left(  \omega,\epsilon\right)   &  =\frac{e^{2\psi-2\gamma}}{8\pi}%
\frac{\omega^{2}\epsilon^{2}}{\rho^{2}+\epsilon^{2}}\left[  AJ_{1}\left(
\omega\sqrt{\rho^{2}+\epsilon^{2}}\right)  \cos\left(  \omega t+\alpha\right)
+BY_{1}\left(  \omega\sqrt{\rho^{2}+\epsilon^{2}}\right)  \cos\left(  \omega
t+\beta\right)  \right]  ^{2}\nonumber\\
&  +\frac{e^{2\psi-2\gamma}}{8\pi}\frac{2\omega^{2}\epsilon^{2}}{\rho
^{2}+\epsilon^{2}}\left[  AJ_{2}\left(  \omega\sqrt{\rho^{2}+\epsilon^{2}%
}\right)  \cos\left(  \omega t+\alpha\right)  +BY_{2}\left(  \omega\sqrt
{\rho^{2}+\epsilon^{2}}\right)  \cos\left(  \omega t+\beta\right)  \right]
\nonumber\\
&  -2\frac{e^{2\psi-2\gamma}}{8\pi}\frac{\omega^{3}\epsilon^{2}}{\sqrt
{\rho^{2}+\epsilon^{2}}}[A^{2}J_{0}\left(  \omega\sqrt{\rho^{2}+\epsilon^{2}%
}\right)  J_{1}\left(  \omega\sqrt{\rho^{2}+\epsilon^{2}}\right)  \cos\left(
2\omega t+2\alpha\right) \nonumber\\
&  +B^{2}Y_{0}\left(  \omega\sqrt{\rho^{2}+\epsilon^{2}}\right)  Y_{1}\left(
\omega\sqrt{\rho^{2}+\epsilon^{2}}\right)  \cos\left(  2\omega t+2\beta
\right)  ]\nonumber\\
&  -2\frac{e^{2\psi-2\gamma}}{8\pi}\frac{\omega^{3}\epsilon^{2}}{\sqrt
{\rho^{2}+\epsilon^{2}}}AB[J_{0}\left(  \omega\sqrt{\rho^{2}+\epsilon^{2}%
}\right)  Y_{1}\left(  \omega\sqrt{\rho^{2}+\epsilon^{2}}\right) \nonumber\\
&  +J_{1}\left(  \omega\sqrt{\rho^{2}+\epsilon^{2}}\right)  Y_{0}\left(
\omega\sqrt{\rho^{2}+\epsilon^{2}}\right)  ]\cos\left(  2\omega t+\alpha
+\beta\right)  . \label{T33.exact}%
\end{align}
The energy-momentum tensor is given by
\begin{equation}
T_{\mu\nu}\left(  \omega\right)  =\lim_{\epsilon\rightarrow0}T_{\mu\nu}\left(
\omega,\epsilon\right)  . \label{Tuv}%
\end{equation}

When $\rho\neq0$, we have
\begin{equation}
T_{\mu\nu}\left(  \omega\right)  =\lim_{\epsilon\rightarrow0}T_{\mu\nu}\left(
\omega,\epsilon\right)  =0. \label{Tuv.rho.1}%
\end{equation}
When $\rho=0$, using
\begin{align}
\lim_{z\rightarrow0}J_{0}\left(  z\right)   &  =1,\text{ }\lim_{z\rightarrow
0}Y_{0}\left(  z\right)  =\frac{2}{\pi}\ln z,\nonumber\\
\lim_{z\rightarrow0}J_{\nu}\left(  z\right)   &  =\frac{1}{\Gamma\left(
\nu+1\right)  }\left(  \frac{z}{2}\right)  ^{\nu},\text{ \ }\nu\neq
0,\nonumber\\
\lim_{z\rightarrow0}Y_{\nu}\left(  z\right)   &  =-\frac{\Gamma\left(
\nu\right)  }{\pi}\left(  \frac{z}{2}\right)  ^{-\nu},\text{ \ \ \ }\nu\neq0,
\end{align}
we have
\begin{align}
T_{00}\left(  \omega\right)   &  =\lim_{\epsilon\rightarrow0}T_{00}\left(
\omega,\epsilon\right)  =T_{11}\left(  \omega,\epsilon\right)  =e^{2\psi
-2\gamma}\frac{B^{2}}{2\pi^{3}}\cos^{2}\left(  \omega t+\beta\right)
\lim_{\epsilon\rightarrow0}\frac{\epsilon^{2}}{\left(  \rho^{2}+\epsilon
^{2}\right)  ^{2}},\nonumber\\
T_{01}\left(  \omega\right)   &  =\lim_{\epsilon\rightarrow0}T_{01}\left(
\omega,\epsilon\right)  =-e^{2\psi-2\gamma}\frac{B^{2}}{2\pi^{3}}\sin\left(
2\omega t+2\beta\right)  \lim_{\epsilon\rightarrow0}\frac{\omega\epsilon^{2}%
}{\rho\left(  \rho^{2}+\epsilon^{2}\right)  }\ln\left(  \omega\sqrt{\rho
^{2}+\epsilon^{2}}\right)  ,\nonumber\\
T_{22}\left(  \omega\right)   &  =\lim_{\epsilon\rightarrow0}T_{22}\left(
\omega,\epsilon\right)  =e^{2\psi-2\gamma}\frac{B^{2}}{2\pi^{3}}\cos
^{2}\left(  \omega t+\beta\right)  \lim_{\epsilon\rightarrow0}\frac
{\epsilon^{2}}{\left(  \rho^{2}+\epsilon^{2}\right)  ^{2}},\nonumber\\
T_{33}\left(  \omega\right)   &  =\lim_{\epsilon\rightarrow0}T_{33}\left(
\omega,\epsilon\right)  =e^{2\psi-2\gamma}\frac{B^{2}}{2\pi^{3}}\left[
\cos^{2}\left(  \omega t+\beta\right)  -\frac{2\pi}{B}\cos\left(  \omega
t+\beta\right)  \right]  \lim_{\epsilon\rightarrow0}\frac{\epsilon^{2}%
}{\left(  \rho^{2}+\epsilon^{2}\right)  ^{2}}.
\end{align}
By use of \cite{ibragimov2009practical}
\[
\lim_{\epsilon\rightarrow0}\frac{\epsilon^{2}}{\left(  x^{2}+y^{2}%
+\epsilon^{2}\right)  ^{2}}=\pi\delta\left(  x\right)  \delta\left(  y\right)
,
\]
we have
\begin{align}
T_{00}\left(  \omega\right)   &  =T_{11}\left(  \omega\right)  =\lim
_{\epsilon\rightarrow0}T_{00}\left(  \omega,\epsilon\right)  =e^{2\psi
-2\gamma}\frac{B^{2}}{2\pi^{2}}\delta\left(  x\right)  \delta\left(  y\right)
\cos^{2}\left(  \omega t+\beta\right)  ,\label{T00.om}\\
T_{22}\left(  \omega\right)   &  =\lim_{\epsilon\rightarrow0}T_{22}\left(
\omega,\epsilon\right)  =e^{2\psi-2\gamma}\frac{B^{2}}{2\pi^{2}}\delta\left(
x\right)  \delta\left(  y\right)  \cos^{2}\left(  \omega t+\beta\right)
,\label{T22.om}\\
T_{33}\left(  \omega\right)   &  =\lim_{\epsilon\rightarrow0}T_{33}\left(
\omega,\epsilon\right)  =e^{2\psi-2\gamma}\frac{B^{2}}{2\pi^{2}}\delta\left(
x\right)  \delta\left(  y\right)  \left[  \cos^{2}\left(  \omega
t+\beta\right)  -\frac{2\pi}{B}\cos\left(  \omega t+\beta\right)  \right]  .
\label{T33.om}%
\end{align}

Because
\[
\lim_{\epsilon\rightarrow0}\frac{T_{01}\left(  \omega,\epsilon\right)
}{T_{00}\left(  \omega,\epsilon\right)  }=0,
\]
$T_{01}\left(  \omega\right)  $ should be regraded as zero:%
\begin{equation}
T_{01}\left(  \omega\right)  =0. \label{T01.om}%
\end{equation}
$T_{01}\left(  \omega\right)  =0$ can be explained from a different way which
we will show below.

Integrating $T_{00}\left(  \omega,\epsilon\right)  $ over the whole space
\begin{align}
E\left(  \omega,\epsilon\right)   &  \equiv\int\sqrt{g}d\rho d\theta d\phi
T_{00}\left(  \omega,\epsilon\right) \nonumber\\
&  =\frac{\omega^{2}\epsilon^{2}L_{z}}{8}[A^{2}\left(  J_{0}^{2}\left(
\omega\epsilon\right)  +J_{1}^{2}\left(  \omega\epsilon\right)  \right)
\cos^{2}\left(  \omega t+\alpha\right)  +B^{2}\left(  Y_{0}^{2}\left(
\omega\epsilon\right)  +Y_{1}^{2}\left(  \omega\epsilon\right)  \right)
\cos^{2}\left(  \omega t+\beta\right) \nonumber\\
&  +2AB\left(  J_{0}\left(  \omega\epsilon\right)  Y_{0}\left(  \omega
\epsilon\right)  +J_{1}\left(  \omega\epsilon\right)  Y_{1}\left(
\omega\epsilon\right)  \right)  \cos\left(  \omega t+\alpha\right)
\cos\left(  \omega t+\beta\right)  ]
\end{align}
and taking the limit $\epsilon\rightarrow0$ give
\begin{equation}
E\left(  \omega\right)  \equiv\lim_{\epsilon\rightarrow0}E\left(
\omega,\epsilon\right)  =\frac{B^{2}L_{z}}{2\pi^{2}}\cos^{2}\left(  \omega
t+\beta\right)
\end{equation}
with $L_{z}\equiv\int_{-\infty}^{\infty}dz$. Again, integrating $T_{01}\left(
\epsilon\right)  $ over the whole space
\begin{align}
P_{\rho}\left(  \omega,\epsilon\right)   &  =\int\sqrt{g}d\rho d\theta d\phi
T_{01}\left(  \omega,\epsilon\right) \nonumber\\
&  =\frac{\omega^{2}\epsilon^{2}L_{z}}{2}[A^{2}\left(  \omega\epsilon
J_{0}^{2}\left(  \omega\epsilon\right)  +\omega\epsilon J_{1}^{2}\left(
\omega\epsilon\right)  -J_{0}\left(  \omega\epsilon\right)  J_{1}\left(
\omega\epsilon\right)  \right)  \sin\left(  \omega t+\alpha\right)
\cos\left(  \omega t+\alpha\right) \nonumber\\
&  +B^{2}\left(  \omega\epsilon Y_{0}^{2}\left(  \omega\epsilon\right)
+\omega\epsilon Y_{1}^{2}\left(  \omega\epsilon\right)  -Y_{0}\left(
\omega\epsilon\right)  Y_{1}\left(  \omega\epsilon\right)  \right)
\sin\left(  \omega t+\beta\right)  \cos\left(  \omega t+\beta\right)
\nonumber\\
&  +AB\left(  \omega\epsilon J_{0}\left(  \omega\epsilon\right)  Y_{0}\left(
\omega\epsilon\right)  +\omega\epsilon J_{1}\left(  \omega\epsilon\right)
Y_{1}\left(  \omega\epsilon\right)  -J_{1}\left(  \omega\epsilon\right)
Y_{0}\left(  \omega\epsilon\right)  \right)  \cos\left(  \omega t+\alpha
\right)  \sin\left(  \omega t+\beta\right) \nonumber\\
&  +AB\left(  \omega\epsilon J_{0}\left(  \omega\epsilon\right)  Y_{0}\left(
\omega\epsilon\right)  +\omega\epsilon J_{1}\left(  \omega\epsilon\right)
Y_{1}\left(  \omega\epsilon\right)  -J_{0}\left(  \omega\epsilon\right)
Y_{1}\left(  \omega\epsilon\right)  \right)  \sin\left(  \omega t+\alpha
\right)  \cos\left(  \omega t+\beta\right)  ]
\end{align}
and taking the limit $\epsilon\rightarrow0$ give
\begin{equation}
P_{\rho}\left(  \omega\right)  \equiv\lim_{\epsilon\rightarrow0}P_{\rho
}\left(  \omega,\epsilon\right)  =\lim_{\epsilon\rightarrow0}\frac{2B^{2}%
L_{z}\omega^{2}}{\pi^{2}}\epsilon\ln\epsilon=0.
\end{equation}
Then the energy-momentum tensor of the metric (\ref{Metric}) with $\psi$ given
by eq. (\ref{single.ps}) and $\gamma$ given by eq. (\ref{single.gamma}) is
\begin{align}
T_{\mu\nu}  &  =e^{2\psi-2\gamma}\frac{B^{2}}{2\pi^{2}}\delta\left(  x\right)
\delta\left(  y\right) \nonumber\\
&  \times\operatorname*{diag}\left(  \cos^{2}\left(  \omega t+\beta\right)
,\cos^{2}\left(  \omega t+\beta\right)  ,\cos^{2}\left(  \omega t+\beta
\right)  ,\cos^{2}\left(  \omega t+\beta\right)  -\frac{2\pi}{B}\cos\left(
\omega t+\beta\right)  \right)  \label{EM}%
\end{align}
with $x=\rho\cos\phi$ and $y=\rho\sin\phi$.

The energy-momentum tensor corresponding to $\psi_{\text{nrad}}$ and
$\gamma_{\text{nrad}}$ in eqs. (\ref{ps.wave}) and (\ref{gamma.wave}) is zero
and the energy-momentum tensor corresponding to $\psi_{\text{rad}}$ and
$\gamma_{\text{rad}}$, eq. (\ref{EM}), is not zero. In other words, the
solutions with $\psi_{\text{rad}}$ and $\gamma_{\text{rad}}$ in eqs.
(\ref{ps.radiation}) and (\ref{gamma.radiation}) have a time-varying energy
density. In general relativity, the total energy itself is the monopole of the
source. The solution with $\psi_{\text{rad}}$ and $\gamma_{\text{rad}}$
describes a gravitational monopole radiation with the time-varying total
energy of the source (\ref{EM}). The solution with $\psi_{\text{nrad}}$ and
$\gamma_{\text{nrad}}$ is a wave solution without sources, which has no
singularity and is just like a plane electromagnetic wave.

There is a problem in the gravitational quadrupole radiation. If insisting
that the energy of the source is invariable in a process of gravitational
radiations, where the energy of the gravitational radiation comes form. If the
energy of the gravitational radiation comes from the source, then the total
energy of the source should decrease with time instead of being invariable. In
general relativity, the charge of the gravitational filed is the energy
density, just like the charge of the electromagnetic field is the electric
charge density. The monopole in general relativity is the integration of the
energy density, i.e., the total energy. The contribution of the gravitational
radiation should be made by the variation of the monopole.

\section{Resonance of cylindrical radiation}

In the following we show that the solution given above contains resonance structures.

To see this, we take
\begin{align}
A\left(  \tau,\lambda\right)   &  =A\delta\left(  \omega_{1}-\lambda\right)
\delta\left(  \tau\right)  ,\nonumber\\
B\left(  \tau,\lambda\right)   &  =B\delta\left(  \omega_{2}-\lambda\right)
\delta\left(  \tau\right)  ,\nonumber\\
\kappa_{1}  &  =\kappa_{2}=\kappa_{3}=0
\end{align}
in the general solution (\ref{ps.general}) and (\ref{gamma.general}):
\begin{align}
\psi &  =A_{1}J_{0}\left(  \omega_{1}\rho\right)  \cos\left(  \omega
_{1}t+\alpha_{1}\right)  +B_{2}Y_{0}\left(  \omega_{2}\rho\right)  \cos\left(
\omega_{2}t+\beta_{2}\right)  ,\label{ps.resonance}\\
\gamma &  =-\frac{A_{1}^{2}}{2}\omega_{1}\rho J_{0}\left(  \omega_{1}%
\rho\right)  J_{1}\left(  \omega_{1}\rho\right)  \cos\left(  2\omega
_{1}t+2\alpha_{2}\right)  -\frac{B_{2}^{2}}{2}\omega_{2}\rho Y_{0}\left(
\omega_{2}\rho\right)  Y_{1}\left(  \omega_{2}\rho\right)  \cos\left(
2\omega_{2}t+2\beta_{2}\right) \nonumber\\
&  -A_{1}B_{2}\frac{\omega_{1}\omega_{2}}{\omega_{1}+\omega_{2}}\rho\left[
J_{0}\left(  \omega_{1}\rho\right)  Y_{1}\left(  \omega_{2}\rho\right)
+Y_{0}\left(  \omega_{2}\rho\right)  J_{1}\left(  \omega_{1}\rho\right)
\right]  \cos\left(  \omega_{1}t+\omega_{2}t+\alpha_{2}+\beta_{2}\right)
\nonumber\\
&  +\frac{A_{1}^{2}}{4}\omega_{1}^{2}\rho^{2}\left[  J_{0}^{2}\left(
\omega_{1}\rho\right)  +2J_{1}^{2}\left(  \omega_{1}\rho\right)  -J_{0}\left(
\omega_{1}\rho\right)  J_{2}\left(  \omega_{1}\rho\right)  \right] \nonumber\\
&  +\frac{B_{2}^{2}}{4}\omega_{2}^{2}\rho^{2}\left[  Y_{0}^{2}\left(
\omega_{2}\rho\right)  +2Y_{1}^{2}\left(  \omega_{2}\rho\right)  -Y_{0}\left(
\omega_{2}\rho\right)  Y_{2}\left(  \omega_{2}\rho\right)  \right] \nonumber\\
&  -A_{1}B_{2}\frac{\omega_{1}\omega_{2}}{\omega_{1}-\omega_{2}}\rho\left[
J_{0}\left(  \omega_{1}\rho\right)  Y_{1}\left(  \omega_{2}\rho\right)
-Y_{0}\left(  \omega_{2}\rho\right)  J_{1}\left(  \omega_{1}\rho\right)
\right]  \cos\left(  \omega_{1}t-\omega_{2}t+\alpha_{1}-\beta_{2}\right)  ,
\label{gamma.resonance}%
\end{align}
where $A_{1}=A\left(  0,\omega_{1}\right)  $, $B_{2}=B\left(  0,\omega
_{2}\right)  $, $\alpha_{1}=\alpha\left(  0,\omega_{1}\right)  $, and
$\beta_{2}=\beta\left(  0,\omega_{2}\right)  $.

The resonance occurs when $\omega_{1}=\omega_{2}$. When $\omega_{1}%
\rightarrow\omega_{2}$, $\gamma\,$diverges due to the existence of the factor
$\frac{1}{\omega_{1}-\omega_{2}}$. Taking $\epsilon=\omega_{1}-\omega
_{2}\rightarrow0$, we have
\begin{equation}
\lim_{\omega_{2}\rightarrow\omega_{1}}\gamma=\lim_{\epsilon\rightarrow0}%
\gamma\sim-\frac{2A_{1}B_{2}}{\pi}\omega_{1}\cos\left(  \alpha_{1}-\beta
_{2}\right)  \lim_{\epsilon\rightarrow0}\frac{1}{\epsilon}.
\end{equation}

The resonance also appears in eq. (\ref{single.gamma}). In eqs.
(\ref{single.ps}) and (\ref{single.gamma}), the radiation part and the
nonradiation part have the same frequency $\omega$. When $A\neq0$ and $B\neq
0$, $\gamma$ in eq. (\ref{single.gamma}) has an aperiodic term being
proportional to $t$. This is also a resonance which means the inpact stores up
and increase with the time $t$.

The solutions (\ref{ps.resonance}) and (\ref{gamma.resonance}) show that the
radiation part (with the frequency $\omega_{2}$) resonates with the
nonradiation part (with the frequency $\omega_{1}$).\textbf{ }

\section{Interaction between radiations}

In this section, we consider the interaction between the radiations. The
theory of gravity is a nonlinear theory, so gravitational radiations in
principle interact with each other.

\subsection{Interaction term in metric}

The factor $\psi$ in the metric (\ref{Metric})\ satisfies a linear equation,
eq. (\ref{Eq.3}), so%
\begin{equation}
\psi=\psi_{1}+\psi_{2} \label{ps.double}%
\end{equation}
\textbf{ }is also a solution when $\psi_{1}$ and $\psi_{2}$ are the solutions
of eq. (\ref{Eq.3}). For convenience, we take $\psi_{1}$ and $\psi_{2}$ given
by eq. (\ref{ps.radiation}) as an example,%
\begin{align}
\psi_{1}  &  =B_{1}Y_{0}\left(  \omega_{1}\rho\right)  \cos\left(  \omega
_{1}t+\beta_{1}\right)  ,\nonumber\\
\psi_{2}  &  =B_{2}Y_{0}\left(  \omega_{2}\rho\right)  \cos\left(  \omega
_{2}t+\beta_{2}\right)  .
\end{align}
These two radiations have the frequencies $\omega_{1}$ and $\omega_{2}$,
respectively. Substituting eq. (\ref{ps.double}) into eqs. (\ref{Eq.1}) and
(\ref{Eq.2}), we have
\begin{equation}
\gamma=\gamma_{1}+\gamma_{2}-\gamma_{\text{int}}, \label{gamma.double}%
\end{equation}
where $\gamma_{1}$ and $\gamma_{2}$ are given by eq. (\ref{gamma.radiation})
with frequencies $\omega_{1}$ and $\omega_{2}$ and
\begin{align}
\gamma_{\text{int}}  &  =B_{1}B_{2}\frac{\omega_{1}\omega_{2}}{\omega
_{1}+\omega_{2}}\rho\left[  Y_{0}\left(  \omega_{1}\rho\right)  Y_{1}\left(
\omega_{2}\rho\right)  +Y_{0}\left(  \omega_{2}\rho\right)  Y_{1}\left(
\omega_{1}\rho\right)  \right]  \cos\left(  \omega_{1}t+\omega_{2}t+\beta
_{1}+\beta_{2}\right) \nonumber\\
&  +B_{1}B_{2}\frac{\omega_{1}\omega_{2}}{\omega_{1}-\omega_{2}}\rho\left[
Y_{0}\left(  \omega_{1}\rho\right)  Y_{1}\left(  \omega_{2}\rho\right)
-Y_{0}\left(  \omega_{2}\rho\right)  Y_{1}\left(  \omega_{1}\rho\right)
\right]  \cos\left(  \omega_{1}t-\omega_{2}t+\beta_{1}-\beta_{2}\right)  .
\label{gamma.int}%
\end{align}
$\gamma_{\text{int}}$ can be understood as an interaction between two
gravitational radiations.

It should be emphasized that in this case when $\omega_{1}=\omega_{2}=\omega$
\begin{align}
\gamma_{\text{int}}  &  =B_{1}B_{2}\omega\rho Y_{0}\left(  \omega\rho\right)
Y\left(  \omega\rho\right)  \cos\left(  2\omega t+\beta_{1}+\beta_{2}\right)
\nonumber\\
&  -\frac{B_{1}B_{2}}{2}\omega^{2}\rho^{2}\left[  Y_{0}^{2}\left(  \omega
\rho\right)  +2Y_{1}^{2}\left(  \omega\rho\right)  -Y_{0}\left(  \omega
\rho\right)  Y_{2}\left(  \omega\rho\right)  \right]  \cos\left(  \beta
_{1}-\beta_{2}\right)
\end{align}
does not diverge. Two radiations do not resonate. The resonance occurs only
between the radiation part and the non-radiation part.

\subsection{Interaction term in energy-momentum tensor}

Now we calculate the energy-momentum tensor of the radiations (\ref{ps.double}%
) and (\ref{gamma.double}).

By the same procedure in section 2, by eqs. (\ref{T00.om}), (\ref{T22.om}),
(\ref{T33.om}), we have
\begin{align}
T_{00}\left(  \omega_{1},\omega_{2}\right)   &  =T_{11}\left(  \omega
_{1},\omega_{2}\right)  =T_{00}\left(  \omega_{1}\right)  +T_{00}\left(
\omega_{2}\right)  +T_{\text{int}}\left(  \omega_{1},\omega_{2}\right)
,\nonumber\\
T_{01}\left(  \omega_{1},\omega_{2}\right)   &  =T_{10}\left(  \omega
_{1},\omega_{2}\right)  =0,\nonumber\\
T_{22}\left(  \omega_{1},\omega_{2}\right)   &  =T_{22}\left(  \omega
_{1}\right)  +T_{22}\left(  \omega_{2}\right)  +T_{\text{int}}\left(
\omega_{1},\omega_{2}\right)  ,\nonumber\\
T_{33}\left(  \omega_{1},\omega_{2}\right)   &  =T_{33}\left(  \omega
_{1}\right)  +T_{33}\left(  \omega_{2}\right)  +T_{\text{int}}\left(
\omega_{1},\omega_{2}\right)  \label{Tuv.int}%
\end{align}
with the interaction term\textbf{ }%
\begin{equation}
T_{\text{int}}\left(  \omega_{1},\omega_{2}\right)  =e^{2\psi-2\gamma}%
\frac{B_{1}B_{2}}{\pi^{2}}\delta\left(  x\right)  \delta\left(  y\right)
\cos\left(  \omega_{1}t+\beta_{1}\right)  \cos\left(  \omega_{2}t+\beta
_{2}\right)  .
\end{equation}

It can be seen that $T_{\mu\nu}\left(  \omega_{1},\omega_{2}\right)  $ which
involves both $\omega_{1}$ and $\omega_{2}$ can be written in three parts:
$T_{\mu\nu}\left(  \omega_{1}\right)  $ which involves only $\omega_{1}$,
$T_{\mu\nu}\left(  \omega_{2}\right)  $ which involves only $\omega_{2}$, and
$T_{\text{int}}\left(  \omega_{1},\omega_{2}\right)  $ which involves both
$\omega_{1}$ and $\omega_{2}$. Here $T_{\text{int}}\left(  \omega_{1}%
,\omega_{2}\right)  $ is an interaction term due to the nonlinearity of the gravity.

Nevertheless, though the nonlinearity of the gravity leads to the existence of
the interaction term $T_{\text{int}}\left(  \omega_{1},\omega_{2}\right)  $,
the time average of $T_{\text{int}}$ vanishes:%
\begin{equation}
\left\langle T_{int}\right\rangle =\int_{0}^{\infty}dtT_{int}\left(
\omega_{1},\omega_{2}\right)  =0.
\end{equation}
The vanishing of the time average of interaction term implies that the result
of a long-time measurement may be linear. That is, when the typical time of a
detector is more larger than the period of a gravitational wave, the source of
radiations observed may be a linear one though the gravitational wave is nonlinear.

As an analogy, we consider a simple example of an energy superposition of two
plane electromagnetic waves. When two plane electromagnetic waves $E_{1}%
=A_{1}\cos\left(  \omega_{1}t+\alpha\right)  $ and $E_{2}=A_{2}\cos\left(
\omega_{2}t+\beta\right)  $ superpose together, the electric field is
\[
E=E_{1}+E_{2}=A_{1}\cos\left(  \omega_{1}t+\alpha\right)  +A_{2}\cos\left(
\omega_{2}t+\beta\right)  ,
\]
where $E$ is the electric field and $A$ is the amplitude. The energy density
of the electric field is
\begin{align}
\varepsilon &  =E^{2}=(E_{1}+E_{2})^{2}\nonumber\\
&  =\varepsilon_{1}+\varepsilon_{2}+\varepsilon_{\text{int}},
\label{elec.energy}%
\end{align}
where $\varepsilon_{1}=A_{1}^{2}\cos^{2}\left(  \omega_{1}t+\alpha\right)  $,
$\varepsilon_{2}=A_{2}^{2}\cos^{2}\left(  \omega_{2}t+\beta\right)  $, and the
interference term
\begin{equation}
\varepsilon_{\text{int}}=2A_{1}A_{2}\cos\left(  \omega_{1}t+\alpha\right)
\cos\left(  \omega_{2}t+\beta\right)  .
\end{equation}
The time average of $\varepsilon_{\text{int}}$ vanishes:
\begin{equation}
\left\langle \varepsilon_{\text{int}}\right\rangle =\int_{0}^{\infty
}dt\varepsilon_{\text{int}}=0.
\end{equation}
The interaction behavior of the energy-momentum tensor (\ref{Tuv.int}) is
similar to the interference behavior of the energy density (\ref{elec.energy}).

\section{Conclusion and outlook}

In linear approximation, the gravitational radiation is a quadrupole
radiation, two radiations superpose linearly and do not interact with each
other. Only the resonance between the radiation and the detector is considered
as the probing scheme of the radiation.

In this paper, we discuss the gravitational radiation based on the exact
cylindrical gravitational wave solutions rather than the linear approximation.
(1) We present a cylindrical gravitational monopole radiation solution which
indicates that the leading contribution of the gravitational radiation may be
the monopole rather than a quadrupole. (2) We consider a new kind of the
resonance between gravitational radiations. Expect the gravity, the radiation
and the spacetime are two separate concepts. Nevertheless, in general
relativity, the concepts of the radiation and the spacetime are mixed up. The
radiation and the spacetime are both described by the metric. We attempt to
separate the radiation and the spacetime in this paper. Following this idea,
we regard the metric without the radiation as a spacetime background. That is,
$\psi_{\text{nrad}}$ and $\gamma_{\text{nrad}}$ in eqs. (\ref{ps.wave}) and
(\ref{gamma.wave}) are regarded as the spacetime background. This idea works
well. The spacetime is also a kind of matter. $\psi_{\text{nrad}}$ and
$\gamma_{\text{nrad}}$ are periodic with the intrinsic frequency $\omega$.
When the gravitational radiation $\psi_{\text{rad}}$ ($\gamma_{\text{rad}}$)
with the same frequency act on the periodic spacetime background, the
resonance occurs. That makes the resonance in the spacetime consistent with
the resonance in Newton mechanics or other physical theories expect the
gravity. We suppose that the resonance between the gravitational radiation and
the spacetime background exists in spite of the symmetry of the system. In
recent years, the gravitational wave detection makes rapid progress. It can be
expected that the resonance between the gravitational radiation and the
spacetime background can be found. (3) We investigate the interaction between
the cylindrical gravitational radiations. The interaction arises both in the
metric and the energy-momentum tensor. Nevertheless, the time average of the
interaction term in the energy-momentum tensor vanishes, which indicates that
the energy-momentum tensor do not interact with each other directly in the
time-averaging level.

In future works, based on gravitational monopole radiation solutions obtained
in this paper, we may figure out if the gravitational dipole radiation
solution exists or not. Besides, the gravitational radiation will lead to the
energy loss of the source. With the conservation law of the energy, we may
define the energy of the cylindrical gravitational radiation in our framework.
We can also consider that the matter wave resonates with the gravitational
radiation based on the preceding work on scattering
\cite{liu2014scattering,li2018scalar,li2019scattering}.

\bigskip

\acknowledgments

We are very indebted to Dr G. Zeitrauman for his encouragement. This work is supported in part by Nankai Zhide foundation and NSF of China under Grant No. 11575125 and No. 11675119.






\end{document}